# Baryon acoustic oscillation, Hubble parameter, and angular size measurement constraints on the Hubble constant, dark energy dynamics, and spatial curvature

Joseph Ryan,[1]⋆ Yun Chen,[2]† and Bharat Ratra,[1]‡
[1]*Department of Physics, Kansas State University, 116 Cardwell Hall, Manhattan, KS 66502, USA*
[2]*Key Laboratory for Computational Astrophysics, National Astronomical Observatories, Chinese Academy of Sciences, Beijing, 100012, China*

**ABSTRACT**

In this paper we use all available baryon acoustic oscillation, Hubble parameter, and quasar angular size data to constrain six dark energy cosmological models, both spatially flat and non-flat. Depending on the model and data combination considered, these data mildly favor closed spatial hypersurfaces (by as much as $1.7\sigma$) and dark energy dynamics (up to a little over $2\sigma$) over a cosmological constant $\Lambda$. The data also favor, at $1.8\sigma$ to $3.4\sigma$, depending on the model and data combination, a lower Hubble constant than what is measured from the local expansion rate.

## 1 INTRODUCTION

The universe is currently undergoing accelerated cosmological expansion. The simplest cosmological model compatible with this acceleration is the standard $\Lambda$CDM model (Peebles 1984), in which the acceleration is powered by a spatially-homogeneous energy density that is constant in time (a cosmological constant $\Lambda$). The standard $\Lambda$CDM model is consistent with many observational constraints (Alam et al. 2017; Farooq et al. 2017; Scolnic et al. 2018; Planck Collaboration 2018) if $\Lambda$ contributes about 70% of the current energy density budget with cold dark matter (CDM) being the next largest contributor, at a little more than 25%.

The standard $\Lambda$CDM model assumes flat spatial hypersurfaces. It has been argued that cosmic microwave background (CMB) anisotropy measurements show that spatial hypersurfaces are very close to being flat, but the recent Planck Collaboration (2016) and Planck Collaboration (2018) CMB anisotropy data analyses in the non-flat case are based on a somewhat arbitrary primordial power spectrum for spatial inhomogeneities. A physically consistent primordial inhomogeneity energy density power spectrum can be generated by inflation, and non-flat inflation models exist which can be used to compute such a power spectrum (for the models, see Gott 1982, Hawking 1984, and Ratra 1985; for the power spectra, see Ratra & Peebles 1995 and Ratra 2017).[1]

When these power spectra (Ratra & Peebles 1995; Ratra 2017) are used in a non-flat $\Lambda$CDM model analysis of CMB anisotropy data (Planck Collaboration 2016) and a large compilation of non-CMB data (Ooba et al. 2018b; Park & Ratra 2018a), a mildly closed $\Lambda$CDM model with ∼1% spatial curvature contribution to the current cosmological energy budget is favored at over $5\sigma$. A similar spatial curvature contribution is favored in dynamical dark energy XCDM and $\phi$CDM models (in which dark energy is modelled as an X-fluid and scalar field, respectively; see Ooba et al. 2018c,d, Park & Ratra 2018b,d). These closed models provide better fits to the low multipole CMB anisotropy data, but the flat models are in better agreement with the higher multipole CMB anisotropy data. The non-flat models are in better agreement with weak lensing measurements, but do a worse job fitting higher redshift cosmic reionization data (Mitra et al. 2018, 2019) and deuterium abundance measurements (Penton et al. 2018).[2]

It has also been found that spatially-flat dynamical dark energy XCDM and $\phi$CDM models provide slightly better overall fits (lower total $\chi^2$) to the current data than does flat $\Lambda$CDM (in the best-fit

---

⋆ E-mail: jwryan@phys.ksu.edu
† E-mail: chenyun@bao.ac.cn
‡ E-mail: ratra@phys.ksu.edu

[1] These non-flat inflation models are slow-roll models, so quantum mechanical fluctuations during inflation in these models result in an untilted

primordial power spectrum. It is possible that these power spectra are too simple, but they are physically consistent; it is not known if the power spectrum used in the Planck non-flat CMB analyses are physically consistent.

[2] Overall the standard tilted flat $\Lambda$CDM model has a lower total $\chi^2$ than the non-flat models, lower by $\Delta\chi^2 \sim$ 10-20, depending on the data compilation and non-flat model used. However, the tilted flat $\Lambda$CDM model is not nested inside any of the three untilted non-flat models, so it is not possible to convert these $\Delta\chi^2$ values to relative goodness-of-fit probabilities (Ooba et al. 2018b,c,d; Park & Ratra 2018b,a,d).

versions of these models the dark energy density has only very mild time dependence; see Ooba et al. 2018a, Park & Ratra 2018b,d, and Sola et al. 2018).[3]

The constraints on spatial curvature and dark energy dynamics discussed above make use of CMB anisotropy data, which requires the assumption of a primordial spatial inhomogeneity power spectrum. As mentioned above, the only currently known physically motivated power spectra in non-flat models are untilted power spectra generated by slow-roll inflation. Such power spectra might not be general enough, so the CMB anisotropy data constraints on spatial curvature derived using these power spectra could be misleading. It is therefore of great importance to constrain spatial curvature and dark energy dynamics using non-CMB data that does not require the assumption of a primordial spatial inhomogeneity power spectrum. For recent studies along these lines, see Farooq et al. (2015), Chen et al. (2016), Yu & Wang (2016), Farooq et al. (2017), Wei & Wu (2017), Rana et al. (2017), Yu et al. (2018), Qi et al. (2018), Ryan et al. (2018), Park & Ratra (2018c), Mukherjee et al. (2019), DES Collaboration (2018a), Zheng et al. (2019), and Ruan et al. (2019).[4]

We recently used Hubble parameter and baryon acoustic oscillation (BAO) measurements to constrain spatial curvature and dark energy dynamics (Ryan et al. 2018).[5] Here we improve upon and extend the analyses of Ryan et al. (2018). Compared to Ryan et al. (2018) we:

- Consider a sixth cosmological model, flat $\Lambda$CDM.
- Update our BAO measurements.
- More accurately compute the size of the sound horizon at the drag epoch for the BAO constraints.
- Treat the Hubble constant $H_0$ as an adjustable parameter to be determined by the data we use.
- Use milliarcsecond quasar angular size versus redshift data (Cao et al. 2017b), alone and in combination with $H(z)$ and BAO data, to constrain cosmological parameters.

We note that, in our analyses here, we make use of the baryon density determined from the Planck 2015 TT + lowP + lensing CMB anisotropy data (Planck Collaboration 2016), as computed in each of the six cosmological models we consider by Park & Ratra (2018b,a,d), in order to calibrate the scale of the BAO sound horizon $r_s$ (which scale is necessary for the computation of distances from BAO data; see below). This means that the constraints we obtain from the BAO data are not completely independent of the Planck 2015 CMB anisotropy data. That said, the baryon density determined from the CMB anisotropy data in the spatially flat models is very consistent with the baryon density determined from deuterium abundance measurements, although it is a little less consistent with these measurements in the non-flat models (Penton et al. 2018).

The new data set that we incorporate in this paper consists of measurements of quasar angular size from Cao et al. (2017b).[6] Measurements of the milliarcsecond-scale angular size of distant radio sources, from data compiled in Gurvits et al. (1999), have been used in the past to constrain cosmological parameters; see Vishwakarma (2001), Lima & Alcaniz (2002), Zhu & Fujimoto (2002), and Chen & Ratra (2003). There is, however, reason to doubt some of these earlier findings. Angular size measurements are only useful if radio sources are standard rulers, as accurate knowledge of the characteristic linear size $l_m$ of the ruler is necessary to convert measurements of the angular size distance into measurements of the angular size, and the estimates of $l_m$ used by Vishwakarma (2001), Lima & Alcaniz (2002), and Zhu & Fujimoto (2002) were inaccurate. To account for the uncertainty in the characteristic linear size $l_m$, Chen & Ratra (2003) marginalized over $l_m$, finding only weak constraints on the cosmological parameters they studied from the angular size data. More recent studies, such as Cao et al. (2017a) and Cao et al. (2017b), based on a sample of 120 intermediate-luminosity quasars recently compiled by Cao et al. (2017b), have more precisely calibrated $l_m$, and these angular size versus redshift data have been used to constrain cosmological parameters (Cao et al. 2017b; Li et al. 2017; Qi et al. 2017; Xu et al. 2018). Here we use these data, in conjunction with $H(z)$ measurements and BAO distance measurements, to constrain cosmological parameters. We find that when the QSO angular size versus redshift data are used in conjunction with the $H(z)$ + BAO data combination, cosmological parameter constraints tighten a bit. We also confirm, as described below, that the QSO data have a large reduced $\chi^2 \sim 3$.

From the full data set, we measure a Hubble constant $H_0$ that is very consistent with the $H_0 = 68 \pm 2.8$ km s$^{-1}$ Mpc$^{-1}$ median statistics estimate (Chen & Ratra 2011a) but is a model-dependent $1.9\sigma$ to $2.5\sigma$ (from the quadrature sum of the error bars) lower than the local expansion rate measurement of $H_0 = 73.48 \pm 1.66$ km s$^{-1}$ Mpc$^{-1}$ (Riess et al. 2018). In the non-flat $\Lambda$CDM model these data are consistent with flat spatial hypersurfaces, while they favor closed geometry at $1.2\sigma$ and $1.7\sigma$ in the non-flat XCDM parametrization and non-flat $\phi$CDM model, respectively. In some of dynamical dark energy models, both flat and non-flat, these data favor dark energy dynamics over a $\Lambda$ (up to a little more $2\sigma$).

In Sec. 2 we describe the models that we study in this paper, Sec. 3 summarizes the data we use, Sec. 4 describes our methods, Sec. 5 describes our results, and we conclude in Sec. 6.

## 2 MODELS

We consider three pairs of dark energy models: the spatially-flat and non-flat models in which dark energy is a cosmological constant $\Lambda$ (flat and non-flat $\Lambda$CDM), a dynamical $X$-fluid with an energy density $\rho_X$ (flat and non-flat XCDM), and a dynamical scalar field $\phi$ (flat and non-flat $\phi$CDM).

In the $\Lambda$CDM models the Hubble parameter as a function of redshift $z$ obeys the Friedmann equation

$$H(z) = H_0\sqrt{\Omega_{m0}(1+z)^3 + \Omega_{k0}(1+z)^2 + \Omega_\Lambda}, \quad (1)$$

where $H_0$ is the Hubble constant, $\Omega_{m0}$ and $\Omega_{k0}$ are the current values of the non-relativistic matter and spatial curvature energy density parameters, and $\Omega_\Lambda$ is the cosmological constant energy density parameter. Conventionally the parameters of the non-flat ΛCDM model are chosen to be $(H_0, \Omega_{m0}, \Omega_\Lambda)$, where $\Omega_{k0} = 1 - \Omega_{m0} - \Omega_\Lambda$, while for the standard spatially-flat ΛCDM model (Peebles 1984) the conventional choice is $(H_0, \Omega_{m0})$ with $\Omega_{k0} = 0$ so $\Omega_\Lambda = 1 - \Omega_{m0}$.

---

[3] For studies of other spatially-flat dynamical dark energy models that fit the data better than does flat $\Lambda$CDM, see Zhang et al. (2017a), Wang et al. (2018), and Zhang et al. (2018).

[4] For possible constraints on spatial curvature from future data, see Witzemann et al. (2018) and Wei (2018).

[5] Hubble parameter data span a large enough redshift range to be able to detect and study the transition from early matter dominated cosmological deceleration to the current dark energy dominated accelerated expansion (see, e.g., Farooq & Ratra 2013; Farooq et al. 2013; Moresco et al. 2016a; Farooq et al. 2017; Jesus et al. 2018; Gómez-Valent 2018). For other uses of Hubble parameter data, see Chen & Ratra (2011b), Chen et al. (2015), Anagnostopoulos & Basilakos (2018), Mamon & Bamba (2018), Geng et al. (2018), and Liu et al. (2018).

[6] For other angular size versus redshift data compilations and constraints, see Daly & Guerra (2002), Podariu et al. (2003), Bonamente et al. (2006), and Chen & Ratra (2012).

The XCDM parametrization is widely used to describe dynamical dark energy. It is physically incomplete because it is based on an ideal, spatially homogeneous $X$-fluid with equation of state relating the pressure and energy density, $p_X = w_X \rho_X$, and negative equation of state parameter $w_X$. This renders it incapable of sensibly describing the evolution of spatial inhomogeneities. When $w_X = -1$, the XCDM parametrization reduces to the physically complete ΛCDM model. In the XCDM parametrization the Hubble parameter is

$$H(z) = H_0 \sqrt{\Omega_{m0}(1+z)^3 + \Omega_{k0}(1+z)^2 + \Omega_{X0}(1+z)^{3(1+w_X)}}, \quad (2)$$

where $\Omega_{X0}$ is the current value of the $X$-fluid energy density parameter. In the non-flat XCDM case the conventional parameters are $(H_0, \Omega_{m0}, \Omega_{k0}, w_X)$, while for flat XCDM, $\Omega_{k0} = 0$, in which case the parameters are $(H_0, \Omega_{m0}, w_X)$.

Dynamical dark energy is modelled as a scalar field, $\phi$, in the physically complete $\phi$CDM model (Peebles & Ratra 1988, Ratra & Peebles 1988, Pavlov et al. 2013).[7] Here the scalar field potential energy density is

$$V = \frac{1}{2}\kappa m_P^2 \phi^{-\alpha}, \quad (3)$$

where $m_P$ is the Planck mass and

$$\kappa = \frac{8}{3}\left(\frac{\alpha+4}{\alpha+2}\right)\left[\frac{2}{3}\alpha(\alpha+2)\right]^{\alpha/2} \quad (4)$$

(we have fixed a typo in Eq. 4 that was present in Ryan et al. 2018). In the $\phi$CDM model $\alpha$ is an adjustable parameter; in the limit $\alpha \to 0$, $\phi$CDM reduces to ΛCDM. The dynamics of the $\phi$CDM model is more complicated than the dynamics of either the ΛCDM model or XCDM parametrization, because the scalar field $\phi$ is a dynamical variable with its own equation of motion. The $\phi$CDM model dynamics is governed by two coupled non-linear ordinary differential equations, the first being the equation of motion for the spatially-homogeneous part of the scalar field

$$\ddot{\phi} + \frac{3\dot{a}}{a}\dot{\phi} - \frac{1}{2}\alpha\kappa m_P^2 \phi^{-\alpha-1} = 0, \quad (5)$$

and the second being the Friedmann equation

$$\left(\frac{\dot{a}}{a}\right)^2 = \frac{8\pi G}{3}(\rho_m + \rho_\phi) - \frac{k}{a^2}. \quad (6)$$

Here the scalar field energy density is

$$\rho_\phi = \frac{1}{2}\dot{\phi}^2 + V, \quad (7)$$

which implies a scalar field energy density parameter

$$\Omega_\phi(z,\alpha) = \frac{8\pi G \rho_\phi}{3H_0^2}, \quad (8)$$

where $\Omega_\phi(z,\alpha)$ is not a simple function of $z$, and must be reconstructed from a numerical solution of eqs. (5) and (6). In the general non-flat case, the Hubble rate can be written

$$H(z) = H_0 \sqrt{\Omega_{m0}(1+z)^3 + \Omega_{k0}(1+z)^2 + \Omega_\phi(z,\alpha)}, \quad (9)$$

and the conventional choice of parameters is $(H_0, \Omega_{m0}, \Omega_{k0}, \alpha)$, while in the spatially-flat case the parameters are $(H_0, \Omega_{m0}, \alpha)$.

## 3 DATA

We use a combination of 120 quasar angular size measurements ("QSO"), 31 expansion rate measurements ("$H(z)$"), and 11 baryon acoustic oscillation measurements ("BAO") to constrain our models. The $H(z)$ data are identical to the data compiled in Ryan et al. (2018) (see that paper for a discussion). The BAO data (see Table 1) have been updated from Ryan et al. (2018); in that paper we used the preprint value of the measurement from Ata et al. (2018), while here we use the published version. Also, we have taken the first six measurements of Table 1 (and the covariance matrix of these measurements) directly from the SDSS website;[8] in Ryan et al. (2018) we did not use the full precision measurements. Additionally, our analysis of the BAO data in this paper differs from that of Ryan et al. (2018), as discussed below.

The BAO measurements collected in Table 1 are expressed in terms of the transverse co-moving distance

$$D_M(z) = \begin{cases} D_C(z) & \text{if } \Omega_{k0} = 0, \\ \frac{c}{H_0\sqrt{\Omega_{k0}}}\sinh\left[\sqrt{\Omega_{k0}}H_0 D_C(z)/c\right] & \text{if } \Omega_{k0} > 0, \\ \frac{c}{H_0\sqrt{|\Omega_{k0}|}}\sin\left[\sqrt{|\Omega_{k0}|}H_0 D_C(z)/c\right] & \text{if } \Omega_{k0} < 0, \end{cases} \quad (10)$$

the Hubble distance

$$D_H(z) = \frac{c}{H(z)}, \quad (11)$$

the volume-averaged angular diameter distance

$$D_V(z) = \left[\frac{cz}{H_0}\frac{D_M^2(z)}{E(z)}\right]^{1/3}, \quad (12)$$

and the line-of-sight comoving distance

$$D_C(z) = \frac{c}{H_0}\int_0^z \frac{dz'}{E(z')}, \quad (13)$$

where $c$ is the speed of light and $E(z) \equiv H(z)/H_0$ (Hogg 1999; Farooq 2013).

All measurements listed in Table 1 are scaled by the size of the sound horizon at the drag epoch $r_s$. This quantity is (see Eisenstein & Hu 1998 for a derivation)

$$r_s = \frac{2}{3k_{eq}}\sqrt{\frac{6}{R_{eq}}}\ln\left[\frac{\sqrt{1+R_d} + \sqrt{R_d + R_{eq}}}{1 + \sqrt{R_{eq}}}\right], \quad (14)$$

where $R_d \equiv R(z_d)$ and $R_{eq} \equiv R(z_{eq})$ are the values of $R$, the ratio of the baryon to photon momentum density

$$R = \frac{3\rho_b}{4\rho_\gamma}, \quad (15)$$

at the drag and matter-radiation equality redshifts $z_d$ and $z_{eq}$, respectively, and $k_{eq}$ is the particle horizon wavenumber at the matter-radiation equality epoch.

---

[7] For discussions of the $\phi$CDM model see Samushia et al. (2007), Yashar et al. (2009), Samushia & Ratra (2010), Samushia et al. (2010), Avsajanishvili et al. (2015), Zhai et al. (2017), Sangwan et al. (2018), Yang et al. (2018), Sola et al. (2018), Tosone et al. (2018), and Singh et al. (2018).

[8] https://sdss3.org/science/boss_publications.php

**Table 1.** BAO data. $D_M\left(r_{s,\mathrm{fid}}/r_s\right)$ and $D_V\left(r_{s,\mathrm{fid}}/r_s\right)$ have units of Mpc, while $H(z)\left(r_s/r_{s,\mathrm{fid}}\right)$ has units of km s$^{-1}$Mpc$^{-1}$ and $r_s$ and $r_{s,\mathrm{fid}}$ have units of Mpc. The uncertainty on the first six measurements is accounted for by the covariance matrix of eq. (20).

| $z$ | Measurement | Value | $\sigma$ | Ref. |
| --- | --- | --- | --- | --- |
| 0.38 | $D_M\left(r_{s,\mathrm{fid}}/r_s\right)$ | 1512.39 | - | Alam et al. (2017) |
| 0.38 | $H(z)\left(r_s/r_{s,\mathrm{fid}}\right)$ | 81.2087 | - | Alam et al. (2017) |
| 0.51 | $D_M\left(r_{s,\mathrm{fid}}/r_s\right)$ | 1975.22 | - | Alam et al. (2017) |
| 0.51 | $H(z)\left(r_s/r_{s,\mathrm{fid}}\right)$ | 90.9029 | - | Alam et al. (2017) |
| 0.61 | $D_M\left(r_{s,\mathrm{fid}}/r_s\right)$ | 2306.68 | - | Alam et al. (2017) |
| 0.61 | $H(z)\left(r_s/r_{s,\mathrm{fid}}\right)$ | 98.9647 | - | Alam et al. (2017) |
| 0.106 | $r_s/D_V$ | 0.336 | 0.015 | Beutler et al. (2011) |
| 0.15 | $D_V\left(r_{s,\mathrm{fid}}/r_s\right)$ | 664 | 25 | Ross et al. (2015) |
| 1.52 | $D_V\left(r_{s,\mathrm{fid}}/r_s\right)$ | 3843 | 147 | Ata et al. (2018) |
| 2.33 | $\frac{(D_H)^{0.7}(D_M)^{0.3}}{r_s}$ | 13.94 | 0.35 | Bautista et al. (2017) |
| 2.36 | $c/(r_s H(z))$ | 9.0 | 0.3 | Font-Ribera et al. (2014) |

To compute $r_s$ as a function of our model parameters, we use the fitting formula presented in Eisenstein & Hu (1998). This calculation also requires $\Omega_b h^2$ as input, and in Ryan et al. (2018) we used $\Omega_b h^2 = 0.02227$ for all models considered. It is more accurate, however, to use the different values of $\Omega_b h^2$ computed by Park & Ratra (2018a,b,d) for each model from the Planck 2015 TT + lowP + lensing CMB anisotropy data (Planck Collaboration 2016), because the values of $\Omega_b h^2$ estimated from CMB anisotropy data are model dependent, and vary significantly between the spatially-flat and non-flat inflation models (Park & Ratra 2018a). The values of $\Omega_b h^2$ that we use are collected in Table 2, without their associated small uncertainties, which we do not account for in our analyses. Here we remind the reader that, because the values of $\Omega_b h^2$ in Table 2 are computed from CMB data, our analysis of the BAO measurements in Table 1 (each of which requires a computation of the sound horizon, which in turn depends on $\Omega_b h^2$) is not completely independent of the CMB data. This should be borne in mind when comparing our $H_0$ measurements to local $H_0$ measurements (see Riess et al. 2018, for example).

Our method of scaling the sound horizon in this paper differs from the method used in Ryan et al. (2018). For studies that scale their measurements by $r_{s,\mathrm{fid}}/r_s$, we use the fitting formula of Eisenstein & Hu (1998) to compute both $r_s$ and $r_{s,\mathrm{fid}}$. For $r_{s,\mathrm{fid}}$ we use the parameters $(\Omega_{m0}, H_0, \Omega_b h^2)$ of the fiducial cosmology adopted in the paper in which the measurement is reported. For measurements scaled only by $r_s$, we again use the fitting formula of Eisenstein & Hu (1998), but we modify it with a multiplicative scaling factor 147.60 Mpc/$r_{s,\mathrm{Planck}}$, where 147.60 Mpc is the value of the sound horizon from Table 4, column 3 of Planck Collaboration (2016), and $r_{s,\mathrm{Planck}}$ is the output of the sound horizon fitting formula from Eisenstein & Hu (1998) when it takes the best-fitting values of $(\Omega_{m0}, H_0, \Omega_b h^2)$ from Planck Collaboration (2016) as input.[9] We do this because the output of the fitting formula in Eisenstein & Hu (1998) deviates by a few per cent from CAMB's output; the scaling factor ensures that $r_s = 147.60$ Mpc when $(\Omega_{m0}, H_0, \Omega_b h^2)$ take their best-fitting values found by Planck Collaboration (2016). We believe that these modifications to the output of the fitting formula result in more accurate determinations of the size of the sound horizon than the scaling employed in Ryan et al. (2018).

---

[9] We thank C.-G. Park for suggesting this.

**Table 2.** Baryon densities for the models we studied.

| Model | $\Omega_b h^2$ | Ref. |
| --- | --- | --- |
| Flat $\Lambda$CDM | 0.02225 | Park & Ratra (2018a) |
| Nonflat $\Lambda$CDM | 0.02305 | Park & Ratra (2018a) |
| Flat XCDM | 0.02229 | Park & Ratra (2018b) |
| Nonflat XCDM | 0.02305 | Park & Ratra (2018b) |
| Flat $\phi$CDM | 0.02221 | Park & Ratra (2018d) |
| Nonflat $\phi$CDM | 0.02303 | Park & Ratra (2018d) |

Recently, Cao et al. (2017a) found that compact structures in intermediate-luminosity radio quasars could serve as standard cosmological rulers. Our QSO data come from a newly compiled sample of these standard rulers from observations of 120 intermediate-luminosity quasars taken over a redshift range of $0.46 < z < 2.76$, with angular sizes $\theta_{\mathrm{obs}}(z)$ and redshifts $z$ listed in Table 1 of Cao et al. (2017b). The corresponding theoretical predictions for the angular sizes can be obtained via

$$\theta_{\mathrm{th}}(z) = \frac{l_m}{D_A(z)}, \qquad (16)$$

where $l_m = 11.03 \pm 0.25$ pc is the intrinsic linear size of the ruler (see Cao et al. 2017b), and

$$D_A(z) = \frac{D_M(z)}{1+z} \qquad (17)$$

is the angular diameter distance at redshift $z$ (see Hogg 1999).

## 4 DATA ANALYSIS METHODS

We use the $\chi^2$ statistic to find the best-fitting parameter values and limits for a given model. Most of the data points we use are uncorrelated, so

$$\chi^2(p) = \sum_{i=1}^{N} \frac{[A_{\mathrm{th}}(p; z_i) - A_{\mathrm{obs}}(z_i)]^2}{\sigma_i^2}. \qquad (18)$$

Here $p$ is the set of model parameters, for example $p = (H_0, \Omega_{m0})$ in the flat $\Lambda$CDM model, $z_i$ is the redshift at which the measured value is $A_{\mathrm{obs}}(z_i)$ with one standard deviation uncertainty $\sigma_i$, and $A_{\mathrm{th}}(p; z_i)$ is the predicted value computed in the model under consideration. The $\chi^2$ expression in eq. (18) holds for the $H(z)$ measurements listed in Table 2 of Ryan et al. (2018) and the BAO measurements listed in the last five lines of Table 1 here.

The measurements in the first six lines of Table 1 are correlated, in which case $\chi^2$ is given by

$$\chi^2(p) = [\mathbf{A}_{\mathrm{th}}(p) - \mathbf{A}_{\mathrm{obs}}]^T \mathbf{C}^{-1} [\mathbf{A}_{\mathrm{th}}(p) - \mathbf{A}_{\mathrm{obs}}], \qquad (19)$$

where $\mathbf{C}^{-1}$ is the inverse of the covariance matrix $\mathbf{C} =$

$$\begin{bmatrix} 624.707 & 23.729 & 325.332 & 8.34963 & 157.386 & 3.57778 \\ 23.729 & 5.60873 & 11.6429 & 2.33996 & 6.39263 & 0.968056 \\ 325.332 & 11.6429 & 905.777 & 29.3392 & 515.271 & 14.1013 \\ 8.34963 & 2.33996 & 29.3392 & 5.42327 & 16.1422 & 2.85334 \\ 157.386 & 6.39263 & 515.271 & 16.1422 & 1375.12 & 40.4327 \\ 3.57778 & 0.968056 & 14.1013 & 2.85334 & 40.4327 & 6.25936 \end{bmatrix}$$

(from the SDSS website).

For the QSO data (Cao et al. 2017b) we use

$$\chi^2(p) = \sum_{i=1}^{N} \left[ \frac{\theta_{\text{th}}(p; z_i) - \theta_{\text{obs}}(z_i)}{\sigma_i + 0.1\theta_{\text{obs}}(z_i)} \right]^2, \quad (21)$$

where $\theta_{\text{th}}(p; z_i)$ is the model-predicted value of the angular size, $\theta_{\text{obs}}(z_i)$ is the measured angular size at redshift $z_i$, and $\sigma_i$ is the uncertainty on the measurement made at redshift $z_i$. The term proportional to $\theta_{\text{obs}}(z_i)$ in the denominator is added to $\sigma_i$ in order to account for systematic uncertainties in the angular size measurements (see the discussion of this point in the first paragraph of Sec. 3 of Cao et al. 2017b).

To determine constraints on the parameters of a given model, we use the likelihood

$$\mathcal{L}(p) = e^{-\chi(p)^2/2}. \quad (22)$$

We are interested in presenting two-dimensional confidence contour plots and one-dimensional likelihoods. To do this, for the models with more than two parameters, we marginalize over the parameters in turn to get one- and two-dimensional likelihoods. In general, we marginalize our likelihood functions by computing integrals of the form

$$\mathcal{L}(p_x) = \int \mathcal{L}(p_x, p_y) \pi(p_y) dp_y, \quad (23)$$

where $p_x$ refers to the set of parameters not marginalized over, $p_y$ refers to the parameter to be marginalized, and $\pi(p_y)$ is a flat, top-hat prior of the form

$$\pi(p_y) = \begin{cases} 1 & \text{if } p_{y,\text{min}} < p_y < p_{y,\text{max}} \\ 0 & \text{otherwise} \end{cases} \quad (24)$$

(see Table 4 for the parameter ranges).[10] For example, in the non-flat $\Lambda$CDM model one of the two-dimensional likelihoods we compute is

$$\mathcal{L}(\Omega_{m0}, \Omega_\Lambda) = \int_{50}^{85} \mathcal{L}(\Omega_{m0}, H_0, \Omega_\Lambda) dH_0, \quad (25)$$

where we integrate the Hubble constant from 50 km s$^{-1}$ Mpc$^{-1}$ to 85 km s$^{-1}$ Mpc$^{-1}$.[11] We then plot the isocontours of $\chi^2(\Omega_{m0}, \Omega_\Lambda) = -2\ln\mathcal{L}(\Omega_{m0}, \Omega_\Lambda)$ in the $\Omega_{m0}$-$\Omega_\Lambda$ subspace of the total parameter space (see Fig. 2).

In addition to plotting the one-dimensional likelihoods for each parameter of each model we consider, we also compute one-sided confidence limits on the best-fitting values of these parameters. The best-fitting value of a parameter $p$ within a given model, after marginalization over the other parameters of the model, is that value $\bar{p}$ which maximizes the one-dimensional likelihood $\mathcal{L}(p)$. To determine the confidence limits $r_n^\pm$ on either side of $\bar{p}$, we compute

$$\frac{\int_{\bar{p}}^{r_n^\pm} \mathcal{L}(p)dp}{\int_{\bar{p}}^{\pm\infty} \mathcal{L}(p)dp} = \sigma_n \quad (26)$$

where $\sigma_{1,2} = 0.6827, 0.9545$ and $r_n^+$, $r_n^-$ are the upper and lower confidence limits out to $\sigma_n$, respectively.

---

[10] We compute the full (not marginalized) likelihoods on a grid. In all flat models and the non-flat $\Lambda$CDM model each parameter has an associated step size of $\Delta p = 0.01$. In the non-flat XCDM parametrization and the non-flat $\phi$CDM model, we use $\Delta H_0 = 0.1$ km s$^{-1}$ Mpc$^{-1}$ to reduce computation time (with all other parameters having $\Delta p = 0.01$).

[11] In Ryan et al. (2018) we used two $H_0$ priors, gaussian with central values and error bars of $\bar{H}_0 \pm \sigma_{H_0} = 68 \pm 2.8$ km s$^{-1}$ Mpc$^{-1}$ (Chen & Ratra 2011a) and $\bar{H}_0 \pm \sigma_{H_0} = 73.24 \pm 1.74$ km s$^{-1}$ Mpc$^{-1}$ (Riess et al. 2016). Here we are instead treating $H_0$ as an adjustable parameter to be determined from the data we use.

In addition to the $\chi^2$ statistic, we also use the Akaike Information Criterion

$$AIC \equiv \chi^2_{\text{min}} + 2k \quad (27)$$

and the Bayes Information Criterion

$$BIC \equiv \chi^2_{\text{min}} + k\ln N \quad (28)$$

(Liddle 2007), where $\chi^2_{\text{min}}$ is the minimum value of $\chi^2$ in the given model, $k$ is the number of parameters in the model, and $N$ is the number of data points. The *AIC* and *BIC* penalize models with a greater number of parameters compared to those with fewer parameters, and as such they can be used to compare the effectiveness of the fits of models with different numbers of parameters.

Although we use Bayesian statistics to analyze our data, this analysis is not complete because we do not compute the Bayesian evidence (a computation which would be prohibitively expensive given that we calculate our likelihoods on a grid rather than using MCMC). Instead we approximate the full Bayesian evidence via $\chi^2$, *AIC*, and *BIC*, which we use to compare our models.

## 5 RESULTS

### 5.1 $H(z)$ + BAO constraints

We discuss our results for the $H(z)$ + BAO data combination first (i.e. without the QSO data). The flat $\Lambda$CDM model (with two free parameters, $H_0$ and $\Omega_{m0}$) two-dimensional $\chi^2$ confidence contours and one-dimensional normalized likelihood curves are plotted in Fig. 1. In Figs. 2-6 we present our results for the non-flat $\Lambda$CDM model, the flat and non-flat XCDM parametrizations, and the flat and non-flat $\phi$CDM models. These results appear in Figs. 1-6 as two-dimensional dashed black likelihood contours and one-dimensional dashed black likelihood curves.

The best-fitting values of the parameters of our models, from their unmarginalized two-, three-, or four-dimensional likelihoods, are listed in Table 3. This table also lists the number of degrees of freedom, $\nu$, and the values of $\chi^2$, *AIC*, and *BIC* that correspond to the best-fitting parameters. The marginalized, one-dimensional best-fitting values of our model parameters, along with their 1$\sigma$ and 2$\sigma$ ranges, are listed in Table 4.

When it is measured using the $H(z)$ + BAO data combination, $\Omega_{m0}$ has consistent best-fitting values, and tight confidence limits, across the models we studied (see Tables 4). For the flat and non-flat $\Lambda$CDM models, $\Omega_{m0} = 0.30^{+0.02}_{-0.01}$ and $\Omega_{m0} = 0.30^{+0.01}_{-0.02}$, respectively. For flat XCDM and flat $\phi$CDM we find $\Omega_{m0} = 0.30^{+0.02}_{-0.01}$, while non-flat XCDM and non-flat $\phi$CDM favor the slightly larger values $\Omega_{m0} = 0.32^{+0.02}_{-0.02}$ and $\Omega_{m0} = 0.31^{+0.02}_{-0.02}$, respectively. Because our $\Omega_{m0}$ step size is 0.01, the 1$\sigma$ error bars on $\Omega_{m0}$ that we list here are probably somewhat inaccurate. The data, however, do determine $\Omega_{m0}$ fairly precisely, with the error bars increasing a bit as the number of model parameters increase, as expected. These $\Omega_{m0}$ estimates are in reasonable agreement with those made by Park & Ratra (2018c) from a similar compilation of $H(z)$ and BAO data.

**Table 3.** Best-fitting parameters of all models

| Model | Data set | $\Omega_{m0}$ | $\Omega_\Lambda$ | $\Omega_{k0}$ | $w_X$ | $\alpha$ | $H_0$[a] | $\nu$ | $\chi^2$ | AIC | BIC |
|---|---|---|---|---|---|---|---|---|---|---|---|
| Flat $\Lambda$CDM | $H(z)$ + BAO | 0.30 | 0.70 | 0 | - | - | 67.99 | 39 | 23.63 | 27.63 | 31.11 |
|  | QSO | 0.32 | 0.68 | 0 | - | - | 68.49 | 117 | 352.05 | 356.05 | 361.63 |
|  | QSO + $H(z)$ + BAO | 0.31 | 0.69 | 0 | - | - | 68.43 | 159 | 376.44 | 380.44 | 386.62 |
| Non-flat $\Lambda$CDM | $H(z)$ + BAO | 0.30 | 0.70 | 0 | - | - | 68.46 | 38 | 23.2 | 29.2 | 34.41 |
|  | QSO | 0.27 | 1 | −0.27 | - | - | 74.62 | 116 | 351.3 | 357.30 | 365.66 |
|  | QSO + $H(z)$ + BAO | 0.30 | 0.73 | −0.03 | - | - | 69.51 | 158 | 375.38 | 381.38 | 390.64 |
| Flat XCDM | $H(z)$ + BAO | 0.30 | - | 0 | −0.94 | - | 66.73 | 38 | 23.29 | 29.29 | 34.50 |
|  | QSO | 0.27 | - | 0 | −1.97 | - | 81.22 | 116 | 351.84 | 357.84 | 366.20 |
|  | QSO + $H(z)$ + BAO | 0.32 | - | 0 | −0.97 | - | 67.90 | 158 | 376.27 | 382.27 | 391.53 |
| Flat $\phi$CDM | $H(z)$ + BAO | 0.30 | - | 0 | - | 0.14 | 66.89 | 38 | 23.41 | 29.41 | 34.62 |
|  | QSO | 0.32 | - | 0 | - | 0.01 | 68.44 | 116 | 352.05 | 358.05 | 366.41 |
|  | QSO + $H(z)$ + BAO | 0.31 | - | 0 | - | 0.07 | 67.94 | 158 | 376.39 | 382.39 | 391.65 |
| Non-flat XCDM | $H(z)$ + BAO | 0.32 | - | −0.23 | −0.73 | - | 66.9 | 37 | 20.94 | 28.94 | 35.89 |
|  | QSO | 0.10 | - | −0.55 | −0.67 | - | 73.9 | 115 | 350.11 | 358.11 | 369.26 |
|  | QSO + $H(z)$ + BAO | 0.31 | - | −0.15 | −0.78 | - | 66.7 | 157 | 372.95 | 380.95 | 393.30 |
| Non-flat $\phi$CDM | $H(z)$ + BAO | 0.31 | - | −0.18 | - | 0.79 | 67.5 | 37 | 21.36 | 29.36 | 36.31 |
|  | QSO | 0.10 | - | −0.43 | - | 2.95 | 72.3 | 115 | 351 | 359 | 370.15 |
|  | QSO + $H(z)$ + BAO | 0.31 | - | −0.14 | - | 0.68 | 67.3 | 157 | 373.49 | 381.49 | 393.84 |

[a] km s$^{-1}$Mpc$^{-1}$.

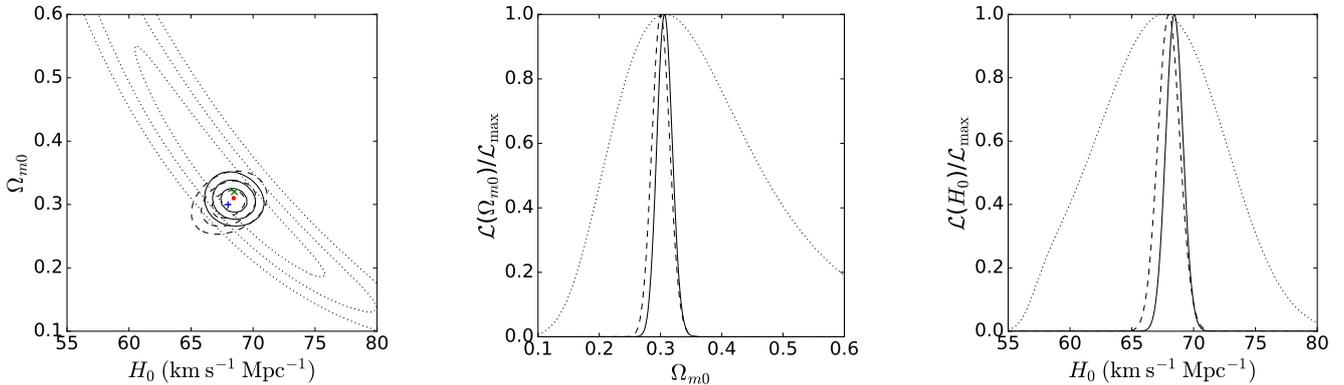

**Figure 1.** Flat $\Lambda$CDM model with QSO, $H(z)$, and BAO data. Left panel: 1, 2, and 3$\sigma$ confidence contours and best-fitting points. Center and right panels: one-dimensional likelihoods for $\Omega_{m0}$ and $H_0$. See text for description and discussion.

The measurements of $H_0$ vary a bit less across the models we studied. For flat (non-flat) $\Lambda$CDM we measure $H_0 = 67.99^{+0.91}_{-0.88}$ $\left(68.24^{+2.39}_{-2.33}\right)$ km s$^{-1}$ Mpc$^{-1}$, while for flat (non-flat) XCDM $H_0 = 66.79^{+2.60}_{-2.32}$ $\left(66.8^{+2.5}_{-2.3}\right)$ km s$^{-1}$ Mpc$^{-1}$, and for flat (non-flat) $\phi$CDM we find $H_0 = 66.13^{+1.38}_{-2.09}$ $\left(67.1^{+2.4}_{-2.3}\right)$ km s$^{-1}$ Mpc$^{-1}$, all with 1$\sigma$ error bars. Our step size is $\Delta H_0 = 0.01$ km s$^{-1}$ Mpc$^{-1}$ for the flat models and non-flat $\Lambda$CDM, which we increased to $\Delta H_0 = 0.1$ km s$^{-1}$ Mpc$^{-1}$ for the non-flat XCDM and $\phi$CDM cases, so the $H_0$ error bars are more accurate than those of the $\Omega_{m0}$ measurements. These six measured $H_0$ values are mutually quite consistent. Aside from the flat $\Lambda$CDM case, the $H_0$ central values and limits are very consistent with those found from a similar $H(z)$ + BAO data compilation in Park & Ratra (2018c). Unlike here where we fix $\Omega_b h^2$ to the values obtained by Park & Ratra (2018a,b,d), Park & Ratra (2018c) allow the baryonic matter density parameter to vary, so the Park & Ratra (2018c) models have an additional free parameter compared to our models; this will have a bigger effect in the flat $\Lambda$CDM case which has the fewest parameters. These $H_0$ measurements are more consistent with the recent median statistics estimate of $H_0 = 68 \pm 2.8$ km s$^{-1}$ Mpc$^{-1}$ (Chen & Ratra 2011a), and with earlier median statistics estimates (Gott et al. 2001, Chen et al. 2003)[12] than with the recent measurement of $H_0 = 73.48 \pm 1.66$ km s$^{-1}$ Mpc$^{-1}$ determined from the local expansion rate (Riess et al. 2018).[13] As a comparison, both our highest and lowest $H_0$ measurements (those of non-flat $\Lambda$CDM and flat $\phi$CDM, respectively) are within $1\sigma$ of the measurement made in Chen & Ratra (2011a), relative to the error bars of that measurement, but they are $1.8\sigma$ (non-flat $\Lambda$CDM) and $3.4\sigma$ (flat $\phi$CDM) lower than the Riess et al. (2018) measurement (here $\sigma$ is the quadrature sum of the two measurement error bars, and these two cases span the range of differences).

---

[12] These $H_0$ measurements are also consistent with many other recent $H_0$ measurements (Chen et al. 2017, Wang et al. 2017, Lin & Ishak 2017, DES Collaboration 2018b, da Silva & Cavalcanti 2018, Gómez-Valent & Amendola 2018, Planck Collaboration 2018, Yu et al. 2018, Zhang 2018, Zhang & Huang 2018, Ruan et al. 2019, Zhang et al. 2019).

[13] We note that other local expansion rate $H_0$ values are slightly lower,

**Table 4.** Best-fitting parameters and $1\sigma$ and $2\sigma$ confidence intervals for all models.

| Model | Data set | Marginalization range[a] | Best-fitting | $1\sigma$ | $2\sigma$ |
|---|---|---|---|---|---|
| Flat $\Lambda$CDM | $H(z)$ + BAO | $0.1 \leq \Omega_{m0} \leq 0.7$<br>$50 \leq H_0 \leq 85$ | $\Omega_{m0} = 0.30$<br>$H_0 = 67.99$ | $0.29 \leq \Omega_{m0} \leq 0.32$<br>$67.11 \leq H_0 \leq 68.90$ | $0.27 \leq \Omega_{m0} \leq 0.33$<br>$66.25 \leq H_0 \leq 68.91$ |
| | QSO + $H(z)$ + BAO | $0.1 \leq \Omega_{m0} \leq 0.7$<br>$50 \leq H_0 \leq 85$ | $\Omega_{m0} = 0.31$<br>$H_0 = 68.44$ | $0.30 \leq \Omega_{m0} \leq 0.32$<br>$67.75 \leq H_0 \leq 69.14$ | $0.28 \leq \Omega_{m0} \leq 0.33$<br>$67.06 \leq H_0 \leq 69.85$ |
| Non-flat $\Lambda$CDM | $H(z)$ + BAO | $0.1 \leq \Omega_{m0} \leq 0.7$<br>$50 \leq H_0 \leq 85$<br>$0.2 \leq \Omega_\Lambda \leq 1$ | $\Omega_{m0} = 0.30$<br>$H_0 = 68.24$<br>$\Omega_\Lambda = 0.70$ | $0.28 \leq \Omega_{m0} \leq 0.31$<br>$65.91 \leq H_0 \leq 70.63$<br>$0.63 \leq \Omega_\Lambda \leq 0.76$ | $0.27 \leq \Omega_{m0} \leq 0.33$<br>$63.60 \leq H_0 \leq 73.03$<br>$0.55 \leq \Omega_\Lambda \leq 0.82$ |
| | QSO + $H(z)$ + BAO | $0.1 \leq \Omega_{m0} \leq 0.7$<br>$50 \leq H_0 \leq 85$<br>$0.2 \leq \Omega_\Lambda \leq 1$ | $\Omega_{m0} = 0.30$<br>$H_0 = 69.32$<br>$\Omega_\Lambda = 0.73$ | $0.29 \leq \Omega_{m0} \leq 0.31$<br>$67.90 \leq H_0 \leq 70.74$<br>$0.67 \leq \Omega_\Lambda \leq 0.78$ | $0.28 \leq \Omega_{m0} \leq 0.33$<br>$66.48 \leq H_0 \leq 72.16$<br>$0.61 \leq \Omega_\Lambda \leq 0.82$ |
| Flat XCDM | $H(z)$ + BAO | $0.1 \leq \Omega_{m0} \leq 0.7$<br>$50 \leq H_0 \leq 85$<br>$-2 \leq w_X \leq 0$ | $\Omega_{m0} = 0.30$<br>$H_0 = 66.79$<br>$w_X = -0.93$ | $0.29 \leq \Omega_{m0} \leq 0.32$<br>$64.47 \leq H_0 \leq 69.39$<br>$-1.05 \leq w_X \leq -0.83$ | $0.27 \leq \Omega_{m0} \leq 0.33$<br>$62.23 \leq H_0 \leq 72.67$<br>$-1.19 \leq w_X \leq -0.74$ |
| | QSO + $H(z)$ + BAO | $0.1 \leq \Omega_{m0} \leq 0.7$<br>$50 \leq H_0 \leq 85$<br>$-2 \leq w_X \leq 0$ | $\Omega_{m0} = 0.31$<br>$H_0 = 68.00$<br>$w_X = -0.97$ | $0.29 \leq \Omega_{m0} \leq 0.32$<br>$66.06 \leq H_0 \leq 70.27$<br>$-1.09 \leq w_X \leq -0.87$ | $0.28 \leq \Omega_{m0} \leq 0.34$<br>$64.22 \leq H_0 \leq 72.67$<br>$-1.22 \leq w_X \leq -0.79$ |
| Flat $\phi$CDM | $H(z)$ + BAO | $0.1 \leq \Omega_{m0} \leq 0.7$<br>$50 \leq H_0 \leq 85$<br>$0.01 \leq \alpha \leq 3$ | $\Omega_{m0} = 0.30$<br>$H_0 = 66.13$<br>$\alpha = 0.15$ | $0.29 \leq \Omega_{m0} \leq 0.32$<br>$64.04 \leq H_0 \leq 67.51$<br>$0.06 \leq \alpha \leq 0.52$ | $0.27 \leq \Omega_{m0} \leq 0.33$<br>$61.95 \leq H_0 \leq 68.73$<br>$0.02 \leq \alpha \leq 0.95$ |
| | QSO + $H(z)$ + BAO | $0.1 \leq \Omega_{m0} \leq 0.7$<br>$50 \leq H_0 \leq 85$<br>$0.01 \leq \alpha \leq 3$ | $\Omega_{m0} = 0.31$<br>$H_0 = 67.19$<br>$\alpha = 0.05$ | $0.30 \leq \Omega_{m0} \leq 0.32$<br>$65.59 \leq H_0 \leq 68.19$<br>$0.02 \leq \alpha \leq 0.36$ | $0.29 \leq \Omega_{m0} \leq 0.34$<br>$63.96 \leq H_0 \leq 69.09$<br>$0.01 \leq \alpha \leq 0.72$ |
| Non-flat XCDM | $H(z)$ + BAO | $0.1 \leq \Omega_{m0} \leq 0.7$<br>$50 \leq H_0 \leq 85$<br>$-2 \leq w_X \leq 0$<br>$-0.7 \leq \Omega_{k0} \leq 0.7$ | $\Omega_{m0} = 0.32$<br>$H_0 = 66.8$<br>$w_X = -0.70$<br>$\Omega_{k0} = -0.15$ | $0.30 \leq \Omega_{m0} \leq 0.34$<br>$64.5 \leq H_0 \leq 69.3$<br>$-0.89 \leq w_X \leq -0.62$<br>$-0.38 \leq \Omega_{k0} \leq 0.01$ | $0.27 \leq \Omega_{m0} \leq 0.36$<br>$62.3 \leq H_0 \leq 71.8$<br>$-1.1 \leq w_X \leq -0.56$<br>$-0.59 \leq \Omega_{k0} \leq 0.14$ |
| | QSO + $H(z)$ + BAO | $0.1 \leq \Omega_{m0} \leq 0.7$<br>$50 \leq H_0 \leq 85$<br>$-2 \leq w_X \leq 0$<br>$-0.7 \leq \Omega_{k0} \leq 0.7$ | $\Omega_{m0} = 0.31$<br>$H_0 = 66.6$<br>$w_X = -0.76$<br>$\Omega_{k0} = -0.12$ | $0.30 \leq \Omega_{m0} \leq 0.33$<br>$64.7 \leq H_0 \leq 68.8$<br>$-0.92 \leq w_X \leq -0.68$<br>$-0.24 \leq \Omega_{k0} \leq -0.02$ | $0.28 \leq \Omega_{m0} \leq 0.34$<br>$62.9 \leq H_0 \leq 71.2$<br>$-1.1 \leq w_X \leq -0.61$<br>$-0.36 \leq \Omega_{k0} \leq 0.07$ |
| Non-flat $\phi$CDM | $H(z)$ + BAO | $0.1 \leq \Omega_{m0} \leq 0.7$<br>$50 \leq H_0 \leq 85$<br>$0.01 \leq \alpha \leq 5$<br>$-0.5 \leq \Omega_{k0} \leq 0.5$ | $\Omega_{m0} = 0.31$<br>$H_0 = 67.1$<br>$\alpha = 0.97$<br>$\Omega_{k0} = -0.2$ | $0.29 \leq \Omega_{m0} \leq 0.33$<br>$64.8 \leq H_0 \leq 69.5$<br>$0.44 \leq \alpha \leq 1.48$<br>$-0.36 \leq \Omega_{k0} \leq -0.06$ | $0.28 \leq \Omega_{m0} \leq 0.34$<br>$62.6 \leq H_0 \leq 71.9$<br>$0.01 \leq \alpha \leq 1.95$<br>$-0.47 \leq \Omega_{k0} \leq 0.05$ |
| | QSO + $H(z)$ + BAO | $0.1 \leq \Omega_{m0} \leq 0.7$<br>$50 \leq H_0 \leq 85$<br>$0.01 \leq \alpha \leq 5$<br>$-0.5 \leq \Omega_{k0} \leq 0.5$ | $\Omega_{m0} = 0.31$<br>$H_0 = 66.8$<br>$\alpha = 0.74$<br>$\Omega_{k0} = -0.15$ | $0.30 \leq \Omega_{m0} \leq 0.32$<br>$65.1 \leq H_0 \leq 68.6$<br>$0.33 \leq \alpha \leq 1.27$<br>$-0.26 \leq \Omega_{k0} \leq -0.06$ | $0.28 \leq \Omega_{m0} \leq 0.34$<br>$63.5 \leq H_0 \leq 70.3$<br>$0.08 \leq \alpha \leq 1.79$<br>$-0.38 \leq \Omega_{k0} \leq 0.02$ |

[a] $H_0$ has units of km s$^{-1}$Mpc$^{-1}$.

As for spatial curvature, we find some evidence in favor of non-flat spatial hypersurfaces, although this evidence is fairly weak.

For non-flat $\Lambda$CDM, we measure $\Omega_{k0} = 0^{+0.06}_{-0.07}$, with $1\sigma$ error bars, consistent with spatial flatness (see Table 4). For the non-flat XCDM parametrization and non-flat $\phi$CDM model, we measure $\Omega_{k0} = -0.15^{+0.16}_{-0.23}$, and $\Omega_{k0} = -0.20^{+0.14}_{-0.16}$, respectively ($1\sigma$ error bars). From these results we can see that non-flat XCDM is consistent with spatial flatness, but non-flat $\phi$CDM favors closed spatial hypersurfaces at a little more than $1.4\sigma$. For these three cases, using a similar $H(z)$ and BAO data compilation, Park & Ratra (2018c) find $\Omega_{k0} = -0.086 \pm 0.078$, $\Omega_{k0} = -0.32 \pm 0.11$, and $\Omega_{k0} = -0.24 \pm 0.15$, respectively, favoring closed geometry at $1.1\sigma$, $2.9\sigma$, and $1.6\sigma$, respectively. Our measurements of spatial curvature are also consistent with the results obtained by other groups, particularly with the model-independent constraints obtained by Yu with slightly larger error bars. See, e.g., Zhang et al. (2017b), Dhawan et al. (2018), and Fernández Arenas et al. (2018).

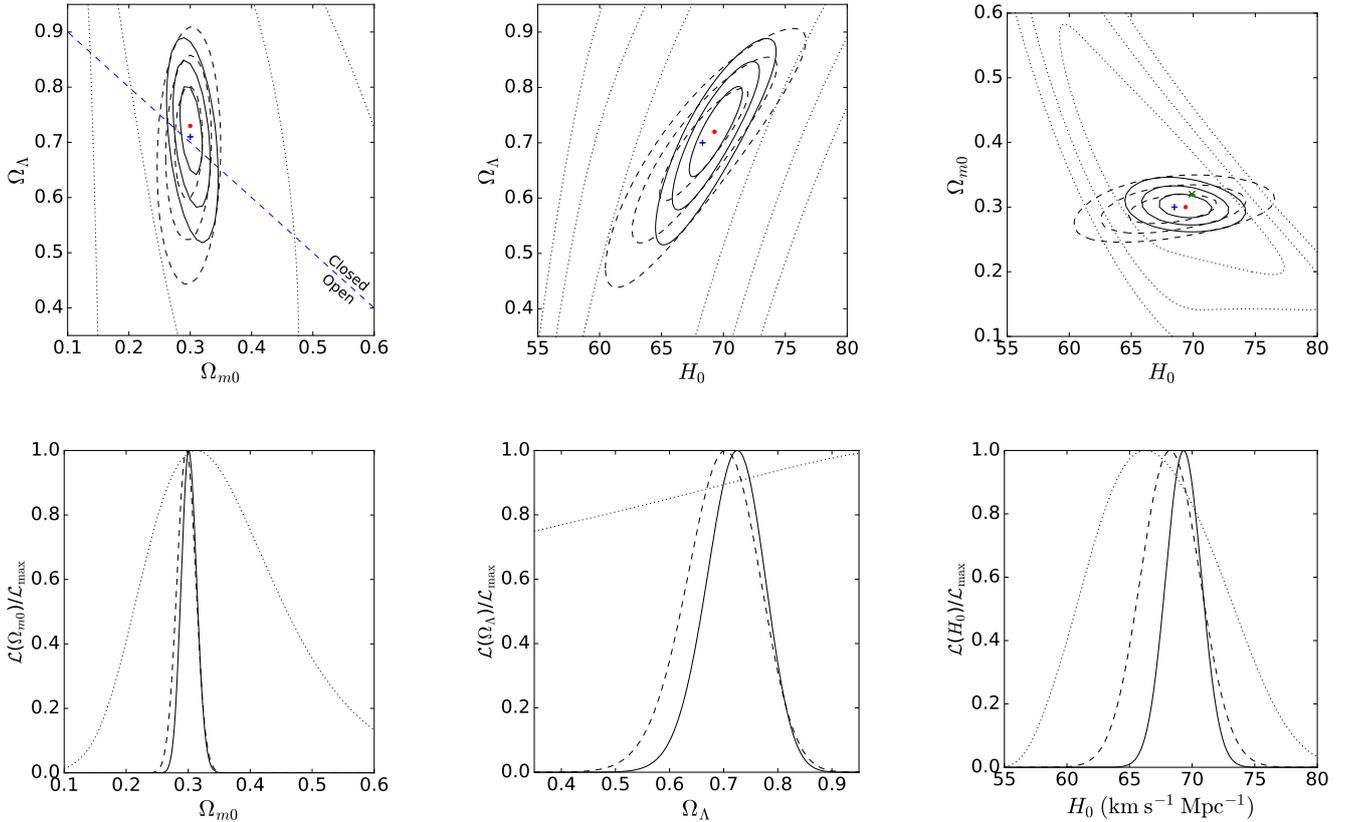

**Figure 2.** Non-flat $\Lambda$CDM model with QSO, $H(z)$, and BAO data. In the top left panel, the blue dashed line demarcates regions of the $\Omega_{m0}$-$\Omega_\Lambda$ parameter space that correspond to spatially open ($\Omega_{k0} > 0$) and spatially closed ($\Omega_{k0} < 0$) models. Points on the line correspond to spatially flat models, with $\Omega_{k0} = 0$. Bottom panels: one-dimensional likelihoods for $\Omega_{m0}$, $\Omega_\Lambda$, and $H_0$. See text for description and discussion.

& Wang (2016), Rana et al. (2017), Wei & Wu (2017), Yu et al. (2018), and Ruan et al. (2019). We find some disagreement in the non-flat $\phi$CDM case with the model-independent studies conducted by Moresco et al. (2016b) and Zheng et al. (2019); when compared to the measurements of $\Omega_{k0}$ made by these groups, we find that our non-flat $\phi$CDM measurement of $\Omega_{k0}$ is not consistent with their measurements to $1\sigma$, although it is consistent to $2\sigma$, owing to the much larger error bars on our measurements.

Our results also show some evidence for dark energy dynamics, although like the evidence for $|\Omega_{k0}| \neq 0$ it is also weak. For example, our measurements in the flat (non-flat) XCDM cases are $w_X = -0.93^{+0.10}_{-0.12}$ at $1\sigma$ ($w_X = -0.70^{+0.08+0.14}_{-0.19-0.40}$ at 1 and $2\sigma$), which both favor quintessence-type dark energy, for which $w_X > -1$, over a $\Lambda$, though to different degrees of statistical significance. The best-fitting value of $w_X$ in the flat XCDM parametrization is within $0.7\sigma$ of $w_X = -1$ (which corresponds to flat $\Lambda$CDM), while the best-fitting value of $w_X$ in the non-flat XCDM parametrization is a little less than $1.6\sigma$ away from $w_X = -1$ (non-flat $\Lambda$CDM in this case). Park & Ratra (2018c) find $w_X = -0.72 \pm 0.16$ ($-0.604 \pm 0.099$) for these two cases, from their $H(z)$ + BAO compilation, which favors quintessence-type dark energy over a $\Lambda$ at $1.8\sigma$ ($4\sigma$). We find marginal evidence for dark energy dynamics in both the flat and non-flat $\phi$CDM models, in which we measure $\alpha = 0.15^{+0.37+0.80}_{-0.09-0.13}$ and $\alpha = 0.97^{+0.51+0.98}_{-0.53-0.96}$, respectively ($1\sigma$ and $2\sigma$ error bars). In both of these cases the measured values of $\alpha$ are a little more than $2\sigma$ away from $\alpha = 0$ (corresponding to $\Lambda$CDM), but this is due to the fact that, in both the flat and non-flat cases, the marginalized likelihood function for $\alpha$ terminates at $\alpha = 0$, the lower limit of our prior range on $\alpha$. The computation of a confidence limit on the low side of the marginalized likelihood function is therefore less meaningful than the computation of a confidence limit on the high side. Our results for $\alpha$ are in less precise agreement with Park & Ratra (2018c) than the results for our other parameters. Park & Ratra (2018c) find for flat $\phi$CDM $\alpha = 2.5 \pm 1.6$ at $1\sigma$ and $\alpha < 6.0$ at $2\sigma$, while for non-flat $\phi$CDM they find $\alpha = 3.1 \pm 1.5$ at $1\sigma$.

Our results here cannot be directly compared to those of our earlier analyses (Ryan et al. 2018), since here $H_0$ is an adjustable parameter to be constrained by the data, while in Ryan et al. (2018) we marginalized over $H_0$, assuming two different gaussian $H_0$ priors. However, we find that the results we have obtained, after marginalizing over $H_0$ with a flat prior, are qualitatively consistent with the results found in Ryan et al. (2018). Further, although we have compared our parameter measurements to those of Park & Ratra (2018c), a direct comparison of our best-fitting $\chi^2$ values to the best-fitting $\chi^2$ values of that paper is not possible because of the different numbers of parameters and data points those authors used,[14] but we agree qualitatively with their result that there are only small differences between the $\chi^2$ of the six models; for each data combination, the six models have relatively similar $\chi^2$, *AIC*, and *BIC* values (see Table 3).

---

[14] Park & Ratra (2018c) use a BAO measurement that we do not; instead of the one gaussian approximation constraint at $z = 2.36$ from Font-Ribera et al. (2014) in Table 1 here, Park & Ratra (2018c) use the probability

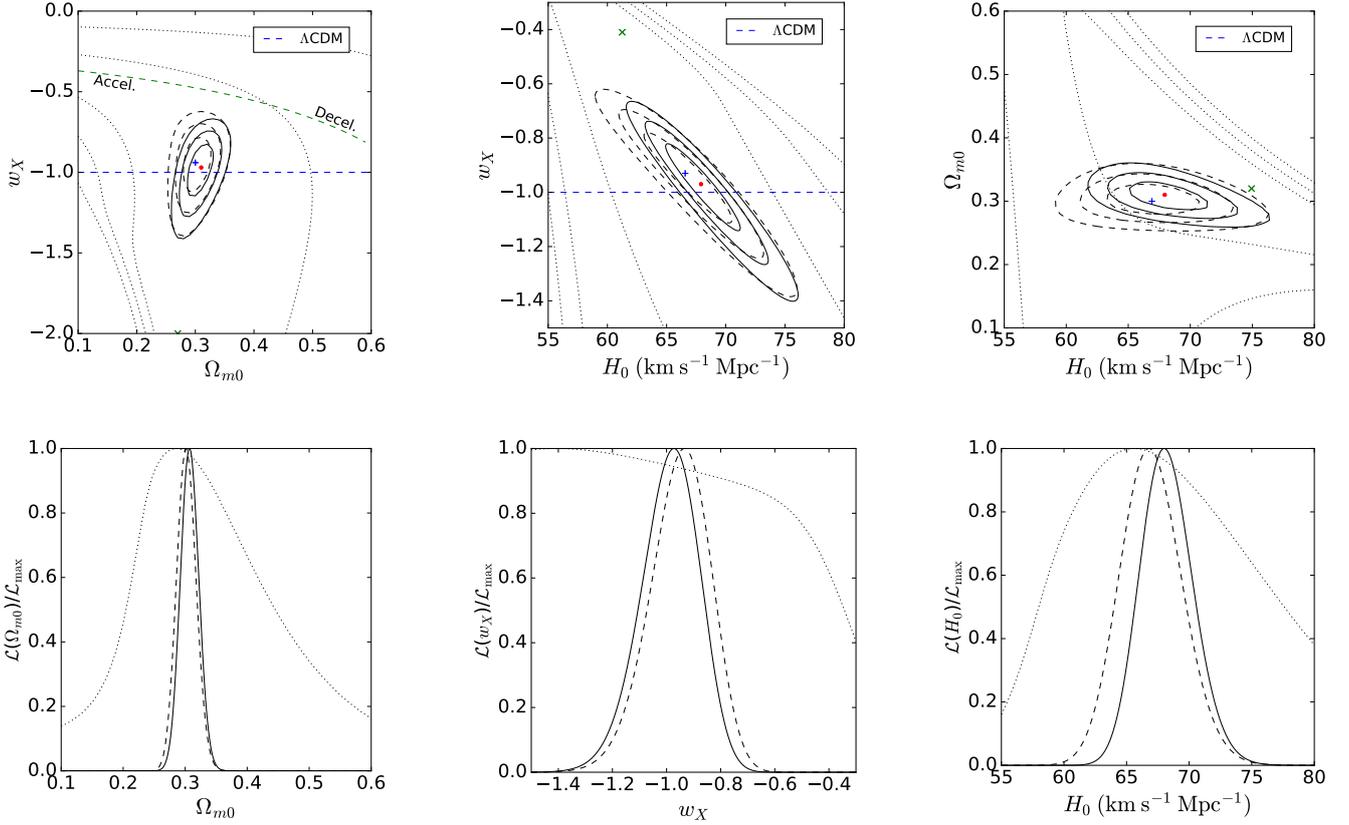

**Figure 3.** Flat XCDM parametrization with QSO, $H(z)$, and BAO data. Top panels: 1, 2, and $3\sigma$ confidence contours and best-fitting points. In the top left and top center panels the horizontal blue dashed line separates quintessence-type parametrizations of dark energy (for which $w_X > -1$) from phantom-type parametrizations of dark energy (for which $w_X < -1$). Points on the blue line (for which $w_X = -1$) correspond to the flat $\Lambda$CDM model. The green dashed curve in the left panel separates models that undergo accelerated expansion now from models that undergo decelerated expansion now. Bottom panels: one-dimensional likelihoods for $\Omega_{m0}$, $w_X$, and $H_0$. See text for description and discussion.

### 5.2 QSO + $H(z)$ + BAO constraints

Our results for the full data set, consisting of QSO data combined with $H(z)$ and BAO data, are presented in Tables 3-4 and in Figs. 1-6. The two-dimensional dotted black likelihood contours and one-dimensional dotted black likelihood curves in Figs. 1-6 correspond to the QSO data alone. The two-dimensional solid black likelihood contours and one-dimensional solid black likelihood curves in Figs. 1-6 correspond to the full data set, namely QSO + $H(z)$ + BAO. By examining the two-dimensional likelihood contours and one-dimensional likelihood curves shown in Figs. 1-6, we see that even though the QSO data by themselves are not able to tightly constrain cosmological parameters, they do contribute to a tightening of the constraints on these parameters when used in combination with $H(z)$ + BAO data.[15]

As with the $H(z)$ + BAO data combination, $\Omega_{m0}$ has consistent central value and error bars when it is measured with the full data set (see Table 4). For the flat and non-flat $\Lambda$CDM models, $\Omega_{m0} = 0.31^{+0.01}_{-0.01}$ and $\Omega_{m0} = 0.30^{+0.01}_{-0.01}$, respectively. For flat XCDM and flat $\phi$CDM we find $\Omega_{m0} = 0.31^{+0.01}_{-0.02}$ and $\Omega_{m0} = 0.31^{+0.01}_{-0.01}$. Non-flat XCDM and non-flat $\phi$CDM have $\Omega_{m0} = 0.31^{+0.02}_{-0.01}$ and $\Omega_{m0} = 0.31^{+0.01}_{-0.01}$, respectively. These measurements have very similar central values and error bars to the measurements made using the $H(z)$ + BAO data combination, as shown by the one-dimensional likelihoods in Figs. 1-6.

For flat (non-flat) $\Lambda$CDM we measure $H_0 = 68.44^{+0.70}_{-0.69}$ $\left(69.32^{+1.42}_{-1.42}\right)$ km s$^{-1}$ Mpc$^{-1}$, while for flat (non-flat) XCDM $H_0 = 68.00^{+2.27}_{-1.94}$ $\left(66.6^{+2.2}_{-1.9}\right)$ km s$^{-1}$ Mpc$^{-1}$, and for flat (non-flat) $\phi$CDM we find $H_0 = 67.19^{+1.00}_{-1.60}$ $\left(66.8^{+1.8}_{-1.7}\right)$ km s$^{-1}$ Mpc$^{-1}$, all $1\sigma$ error bars. Compared to the cases without the QSO data in Sec. 5.1, the central $H_0$ values here are a little larger (except in the non-flat XCDM and $\phi$CDM cases) and the error bars are a little smaller. These $H_0$ estimates are still in very good agreement with that from median statistics (Chen & Ratra 2011a) but differ

---

distribution that describes the shift of the BAO peak position in both the perpendicular and parallel directions to the line of sight.
[15] We confirm the high reduced $\chi^2$ values for the QSO angular size data (see Tables 3-4) found earlier by Zheng et al. (2017) (see Table 2 of that paper), Qi et al. (2017) (see Table 5 of that paper), and Xu et al. (2018) (see Table 2 of that paper). What causes this is apparently not yet understood.

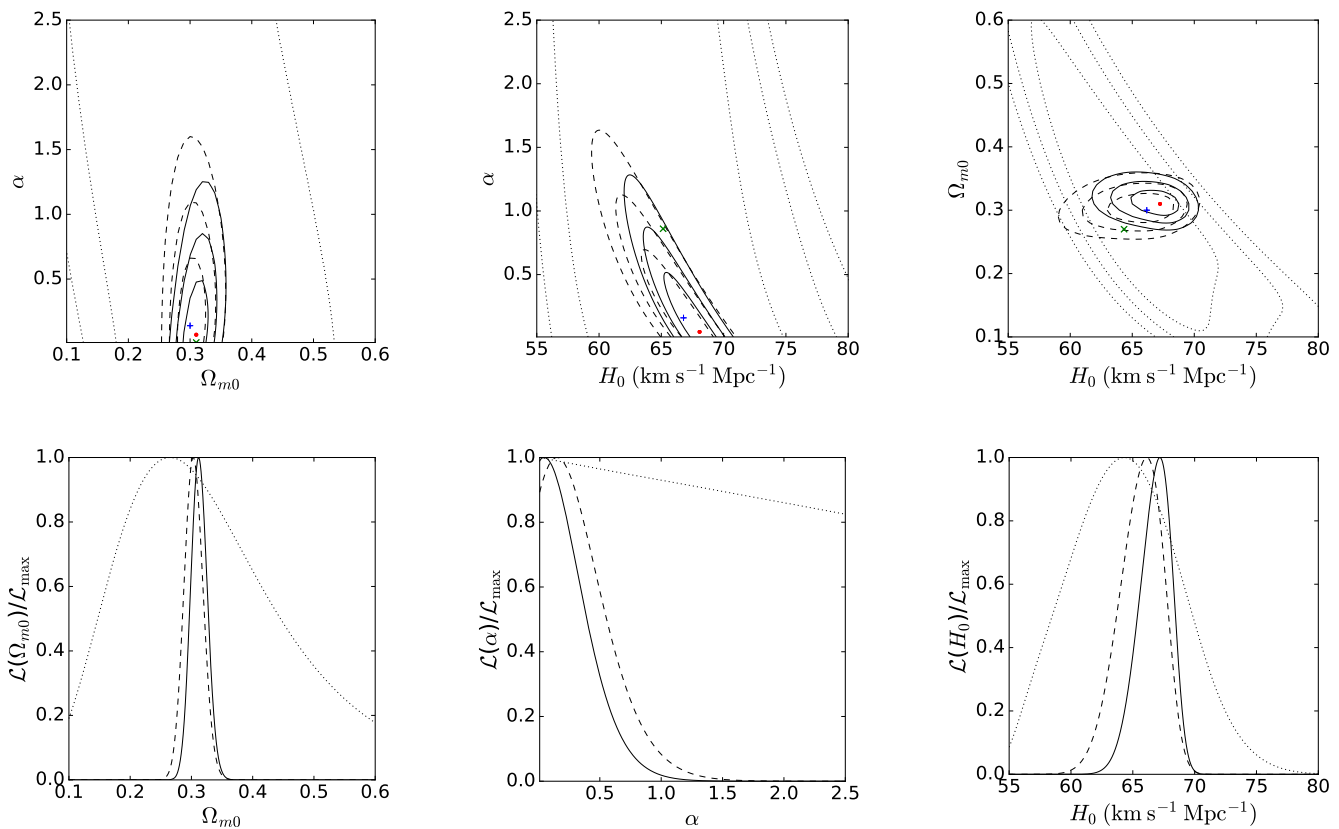

**Figure 4.** Flat $\phi$CDM model with QSO, $H(z)$, and BAO data. Top panels: 1, 2, and 3$\sigma$ confidence contours and best-fitting points. Points on the $\alpha = 0$ line in the top left and top center panels correspond to the flat $\Lambda$CDM model. Bottom panels: one-dimensional likelihoods of $\Omega_{m0}$, $\alpha$, and $H_0$. See text for description and discussion.

from that measured from the local expansion rate (Riess et al. 2018), being between 1.9$\sigma$ (non-flat $\Lambda$CDM) and 2.5$\sigma$ (non-flat XCDM) lower (as before, $\sigma$ refers to the quadrature sum of the error bars on the two measurements).

When we measure the curvature energy density parameter using the full data set, we find in the non-flat $\Lambda$CDM model that $\Omega_{k0} = -0.03^{+0.05}_{-0.06}$, which is again consistent with flat spatial hypersurfaces, but with slightly tighter error bars. The same pattern holds when we measure $\Omega_{k0}$ in the non-flat XCDM parametrization and the non-flat $\phi$CDM model, in which $\Omega_{k0} = -0.12^{+0.10}_{-0.12}$ and $\Omega_{k0} = -0.15^{+0.09}_{-0.11}$, respectively. Both of these measurements are slightly more consistent with closed spatial hypersurfaces than the corresponding measurements made using only the $H(z)$ + BAO data combination, being 1.2$\sigma$ (non-flat XCDM) and 1.7$\sigma$ (non-flat $\phi$CDM) away from spatial flatness.

The parameters that govern dark energy dynamics move closer to $\Lambda$CDM when we measure them with the full data set. In the flat (non-flat) XCDM parametrization, $w_X = -0.97^{+0.10}_{-0.12}$ $\left(w_X = -0.76^{+0.08}_{-0.16}\right)$, with 1$\sigma$ error bars. In both cases we find that the addition of QSO data to the $H(z)$ + BAO data drives the value of $w_X$ closer to $w_X = -1$, the value that it takes in the flat and non-flat $\Lambda$CDM models (although $w_X$ is still over 1$\sigma$ larger than $-1$ in the non-flat case). Something similar happens to $\alpha$; in the flat (non-flat) $\phi$CDM model we measure $\alpha = 0.05^{+0.31+0.68}_{-0.03-0.04}$ $\left(\alpha = 0.74^{+0.53+1.05}_{-0.41-0.66}\right)$, with 1 and 2$\sigma$ error bars that are tighter in the flat case than they are when $\alpha$ is measured using only $H(z)$ + BAO data. As in XCDM, the parameter that controls the dark energy dynamics, $\alpha$, is driven closer to $\alpha = 0$, the value that it takes when $\phi$CDM reduces to $\Lambda$CDM (though $\alpha$ is still measured to be about 2$\sigma$ away from zero in the non-flat case).

## 6 CONCLUSION

We analyzed a total of 162 observations, 120 of which were measurements of the QSO angular sizes from Cao et al. (2017b), with the remaining 42 measurements being a combination of $H(z)$ data and distance measurements from baryon acoustic oscillations (compiled in Ryan et al. 2018).

Our methods and models were largely the same in this paper as in Ryan et al. (2018), with a few key differences. First, we treated $H_0$ as a free parameter, so as to obtain constraints on its value within the models we studied. We also presented results for each of our data sets separately and in combination (in Ryan et al. 2018 we only presented results for the $H(z)$ + BAO data combination), and treated the sound horizon and $D_M$-$H(z)$ covariance matrix more accurately (see Sec. 4). After accounting for these differences, we find that our results for the energy density parameters $\Omega_{m0}$ and $\Omega_\Lambda$, the dark energy equation of state $w_X$, and the $\phi$CDM potential energy density parameter $\alpha$ are largely consistent with those of Ryan et al. (2018).

Adding QSO data to $H(z)$ and BAO data tightens parameter constraints in some of the models we studied. In particular, using the full data set, we find that there is some evidence for closed spatial hypersurfaces in dynamical dark energy models, but that this evidence is only marginally significant (being between $1.2\sigma$ and $1.7\sigma$, depending on the model considered). We also find that there is marginal evidence for dark energy dynamics in both flat and non-flat models, ranging from around $0.7\sigma$ to a little more than $2\sigma$, depending on the model. A little more significant is the evidence we find in favor of a lower value of the Hubble constant. Our $H_0$ results are more consistent with the results of Chen & Ratra (2011a) and Planck Collaboration (2018) than that of Riess et al. (2018), being between $1.9\sigma$ lower than the measurement made by Riess et al. (2018) in the non-flat $\Lambda$CDM model and $2.5\sigma$ lower than said measurement in the non-flat XCDM parametrization (although these error bars on $H_0$ are not as wide as the error bars $H_0$ when $H_0$ is measured using only the $H(z)$ + BAO data combination).

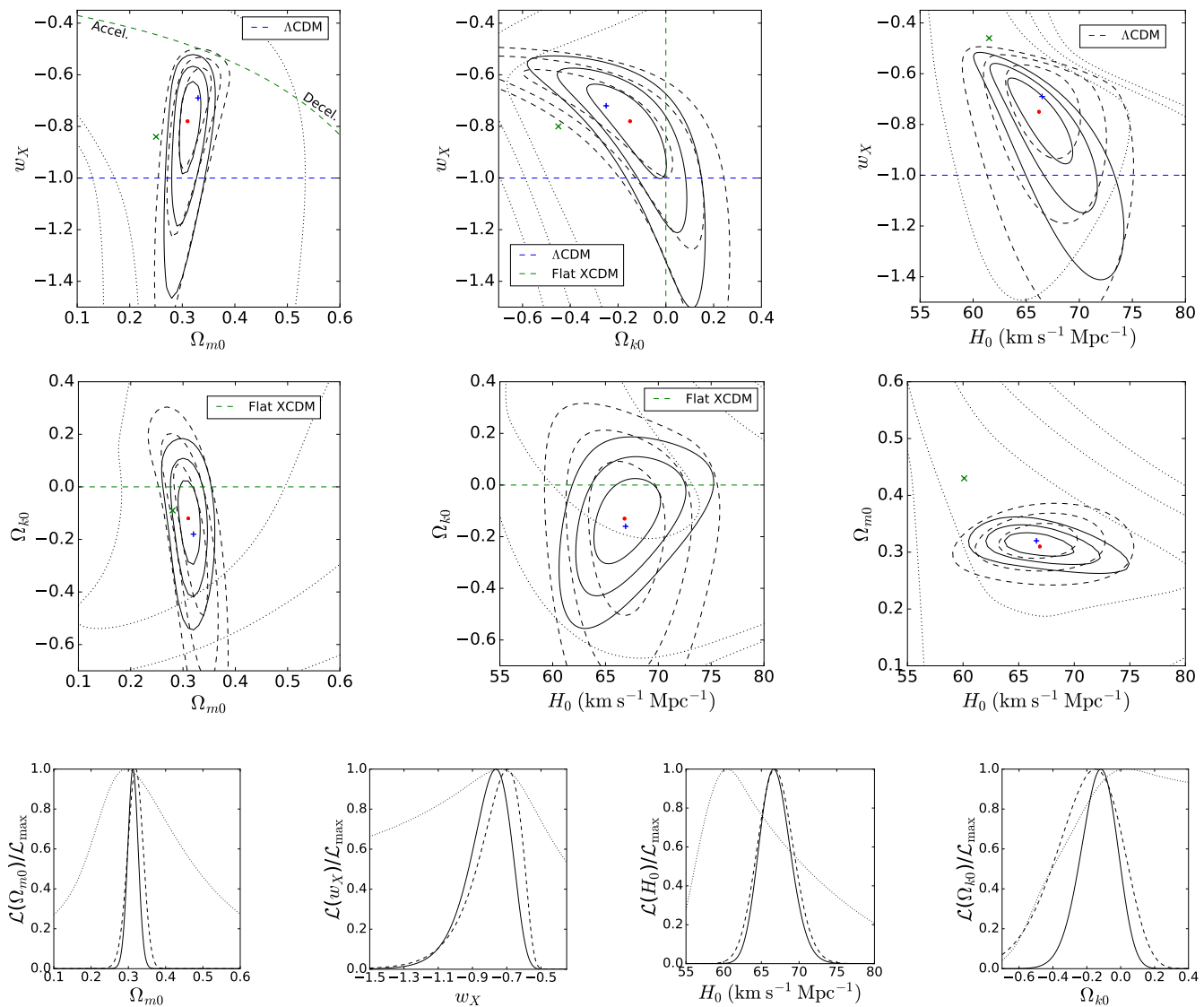

**Figure 5.** Non-flat XCDM parametrization with QSO, $H(z)$ and BAO data. Top and middle row: 1, 2, and $3\sigma$ confidence contours and best-fitting points. In the top panels, the horizontal blue dashed line separates quintessence-type parametrizations of dark energy (for which $w_X > -1$) from phantom-type parametrization of dark energy (for which $w_X < -1$). Points on the blue line (for which $w_X = -1$) correspond to the non-flat $\Lambda$CDM model. The green dashed curve in the top left panel separates models that undergo accelerated expansion now from models that undergo decelerated expansion now. The vertical green dashed line in the top center panel, and the horizontal green dashed lines in the left and center panels of the middle row, separate spatially closed models (for which $\Omega_{k0} < 0$) from spatially open models (for which $\Omega_{k0} > 0$). Bottom panels: one-dimensional likelihoods for $\Omega_{m0}$, $w_X$, $H_0$, and $\Omega_{k0}$. See text for description and discussion.

## ACKNOWLEDGEMENTS

We thank Chan-Gyung Park and Lado Samushia for helpful discussions, we thank Dave Turner for providing valuable technical advice, and we thank Shulei Cao for pointing out an error in an earlier version of this paper. Some of the computing for this project was performed on the Beocat Research Cluster at Kansas State University, which is funded in part by NSF grants CNS-1006860, EPS-1006860, and EPS-0919443. This work was partially funded by DOE Grant DE-SC0019038. YC received support from the National Natural Science Foundation of China (Nos. 11703034, 11773032 and 11573031), and the NAOC Nebula Talents Program.

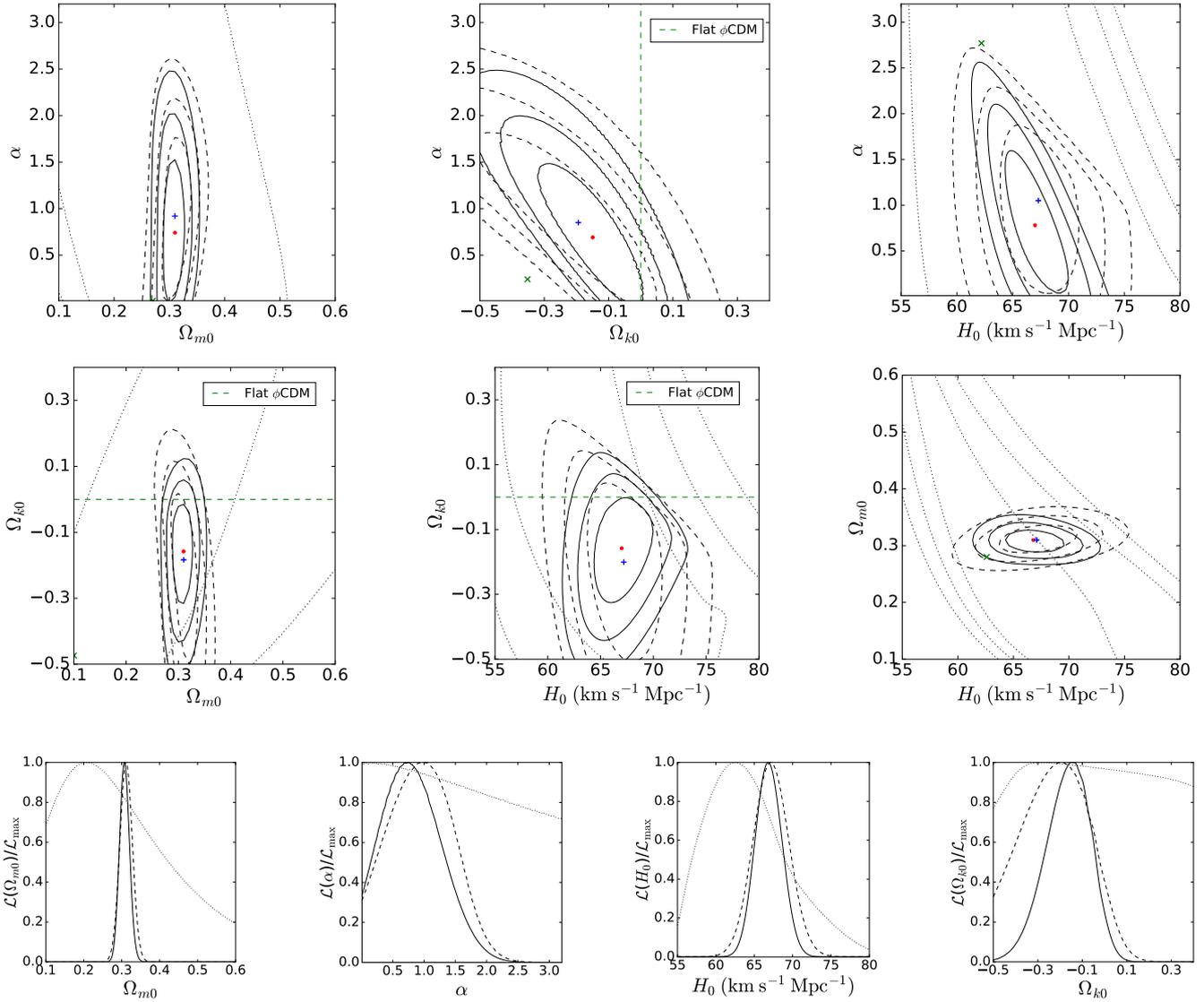

**Figure 6.** Non-flat $\phi$CDM model with QSO, $H(z)$, and BAO data. Top and middle rows: 1, 2, and 3$\sigma$ confidence contours and best-fitting points. The vertical green dashed line in the top center panel, and the horizontal green dashed lines in the middle left and middle center panels, separate spatially closed models (with $\Omega_{k0} < 0$) from spatially open models (with $\Omega_{k0} > 0$). Points on the $\alpha = 0$ line in the top panels correspond to the non-flat $\Lambda$CDM model. Bottom row: one-dimensional likelihoods for $\Omega_{m0}$, $\alpha$, $\Omega_{k0}$, and $H_0$. See text for description and discussion.

This paper has been typeset from a TeX/LaTeX file prepared by the author.